\documentclass[submission,copyright,creativecommons]{eptcs}
\providecommand{\event}{QSQW 2020} 

\usepackage{amsmath}
\usepackage{mathtools, nccmath}
\usepackage{eucal}
\usepackage{float}
\usepackage{amssymb}
\usepackage{upquote}
\usepackage{hyperref}
\usepackage{cleveref}
\usepackage{amsfonts}

\usepackage{graphicx}
\graphicspath{ {./figures/} }
\usepackage{float}
\usepackage{subfigure}
\usepackage{cite}
\usepackage{xcolor}

\newcommand{\ket}[1]{|#1\rangle}

\newcommand{\bra}[1]{\langle #1|}

\DeclareMathAlphabet{\pazocal}{OMS}{zplm}{m}{n}

\title{A Generalized Quantum Optical Scheme for Implementing Open Quantum Walks}
\author{Ayanda Romanis Zungu
\institute{Centre for Space Research \\ North-West University\\
Mahikeng, 2745, South Africa}
\email{arzngu@gmail.com}
\and
IIya Sinayskiy$^{\dagger}$ \qquad Francesco Petruccione$^{\ddagger}$ 
\institute{Quantum Research Group, School of Chemistry and Physics\\ University of KwaZulu-Natal\\
Durban, 4001, South Africa}
\email{\quad $^{\dagger}$sinayskiy@ukzn.ac.za \qquad $^{\ddagger}$petruccione@ukzn.ac.za}
}
\def\titlerunning{A Generalized Quantum Optical Scheme for Implementing Open Quantum Walks}
\def\authorrunning{A.R. Zungu, I. Sinayskiy \& F. Petruccione}
\begin{document}

\maketitle

\begin{abstract}
{\bf Abstract.} Open quantum walks (OQWs) are a new type of quantum walks which are entirely driven by the dissipative interaction with external environments and are formulated as completely positive trace-preserving maps on graphs. A generalized quantum optical scheme for implementing OQWs that includes non-zero temperature of the environment is suggested. In the proposed quantum optical scheme, a two-level atom plays the role of the {\ttfamily"}walker{\ttfamily"}, and the Fock states of the cavity mode correspond to the lattice sites for the {\ttfamily"}walker{\ttfamily"}. Using the small unitary rotations approach the effective dynamics of the system is shown to be an OQW. For the chosen set of parameters, an increase in the temperature of the environment causes the system to reach the asymptotic distribution much faster compared to the scheme proposed earlier where the temperature of the environment is zero. For this case the asymptotic distribution is given by a steady Gaussian distribution.
\end{abstract}

\section{Introduction}
The complete description of any realistic quantum system includes the unavoidable effects of the interaction with the environment \cite{breuer2002theory}. Such systems are characterized by the presence of decoherence and dissipation and are treated as an ``open'' quantum system. The influence of interaction with an environment plays a fundamental role in moving from the quantum to the classical domain. However, it has been reported that the interaction with the environment can not only create complex entangled states \cite{diehl2008quantum,kraus2008preparation,kastoryano2011dissipative}, but also allows for universal quantum computation \cite{verstraete2009quantum}. One of the well established approaches to formulate quantum computing algorithms is in the language of quantum walks (QWs) \cite{kempe2003quantum,aharonov1993quantum}, the quantum version of classical random walks (CRWs). QWs were introduced in two forms, namely, continuous \cite{aharonov1993quantum} and discrete \cite{kempe2003quantum} in time, and can be used to perform universal quantum computation \cite{childs2009universal}. QWs have also been used to analyze energy transport in biological systems \cite{mohseni2008environment}.

Remarkably, taking into account decoherence and dissipation effects in QWs reduces its applicability for quantum computation \cite{kendon2007decoherence}. Although in very small amounts decoherence has been found to be useful; hence, it is therefore worthwhile to come up with a framework that includes these environmental effects as an ingredient. This framework will be based on the non-unitary dynamics induced by the environment, perhaps, this will lead to new interesting quantum behaviors \cite{attal2012oopen, sinayskiy2012efficiency}. 

More recently, a new type of QWs called open QWs (OQWs) were proposed by Attal et al. \cite{attal2012open, attal2012oopen} with the aim of incorporating dissipation and decoherence effects \cite{breuer2002theory}. OQWs are formulated as a quantum Markov chain on lattices or graphs. Unlike unitary quantum walks (UQWs) \cite{kempe2003quantum,childs2009universal} where dissipation and decoherence effects need to be minimised or eliminated when dealing with quantum systems \cite{kendon2007decoherence,ellinas2007quantum}, in OQWs, these effects are naturally included into the description of the dynamics of the open quantum ``walker''. OQWs deal with density matrices instead of pure states. Mathematically they are formulated as completely positive trace-preserving maps (CPTP maps) on graphs \cite{breuer2002theory, kraus1983states}. In OQWs the transition between the nodes is driven purely by the dissipative interaction with local environments and the probability to find the quantum walker on a particular node depends not only on the structure of the underlying graph, but also on the inner state of the walker. Also, OQWs are not exploiting the interference between different positions of the quantum walker as do UQWs \cite{kempe2003quantum}.

The framework of OQWs appears to be very useful and can be used to perform efficient dissipative quantum computation and state engineering \cite{attal2012oopen, sinayskiy2012efficiency}. For recent developments on OQWs the reader is referred to \cite{sinayskiy2019open} and references cited therein. In particular \cite{attal2015central, konno2013limit, sadowski2016central} have shown that the asymptotic behavior of OQWs leads to a central limit theorem. And in general, for large times, the position probability distribution of OQWs converges to Gaussian distributions \cite{attal2015central, konno2013limit}.

Not so long ago, Sinayskiy and Petruccione suggested two possible approaches to implement OQWs: first, they suggested a quantum optics implementation of OQWs by using an effective operator formalism \cite{sinayskiy2014quantum}, second, they followed the traditional theory of open quantum systems and derived OQWs based on the microscopic system-bath setup \cite{sinayskiy2013microscopic,sinayskiy2015microscopic}. In the quantum optical implementation of OQWs,  \cite{sinayskiy2014quantum} used an example of the two-level system in the cavity in the dispersive regime. This was done by considering a composite quantum optical system. For this case, an interaction between a two-level atom with a quantized mode of the electromagnetic field at zero temperature. More precisely, \cite{sinayskiy2014quantum} considered  a particular system in a regime in which the OQW formalism becomes a natural way of describing the system effective dynamics.

The state of the two-level system corresponds to an inner degree of freedom of the ``walker'', and the Fock states correspond to the lattice sites for the ``walker''. By using the method of small unitary rotations \cite{klimov2000method,Klimov_2002,klimov2009group}, the effective dynamics of the system were shown to be an OQW. The only dissipative process considered for obtaining an OQW was the spontaneous emission in the system. Although this scheme leads to OQW, the dynamics of the walker is relatively poor in comparison to traditional microscopic approaches \cite{sinayskiy2015microscopic}. The aim of this paper is to generalize the  simple case which was developed earlier by \cite{sinayskiy2014quantum} to include external driving of the atom and non-zero temperature of the environment.

This paper is structured as follows: In Sec. \ref{Formalism}, we briefly review the formalism of OQWs. In Sec. \ref{model}, we apply the method of the small unitary rotations to the quantum master equation describing the dynamics of the system. After that, we obtain a generalized master equation. The time discretization of the generalized master equation leads to OQW formalism.  Finally, in Sec. \ref{Conclusion}, we conclude.

\section{Formalism} \label{Formalism}

OQWs are defined as quantum walks on graphs \cite{attal2012open, attal2012oopen}, where the transitions between the nodes are driven by the dissipative interaction with an environment. This process is realized through repeated application of a specific completely positive trace-preserving (CPTP) map. In order to describe OQWs, we consider a walk on a graph with the set of vertices $\mathcal{V}$ and directed edges $\{(i, j): i, j \in \mathcal{V}\}$. Here the set $\mathcal{V}$ of vertices may be finite or countably infinite. The dynamics on the graph are described by the space of states $\mathcal{K}=\mathbb{C}^\mathcal{V}$ with orthonormal basis $\{\ket{i}\}_{i\in \mathcal{V}}$ indexed by $\mathcal{V}$. We describe an internal degree of freedom of the walker e.g., spin, polarization or $n$-energy levels, by attaching a separable Hilbert space $\mathcal{H}_S$ to each node of the graph, such that any state of the walker is described by a density matrix $\rho$ on the directed product of the Hilbert spaces $\mathcal{H}_S\otimes \mathcal{K}$.

\begin{figure}[H]
\centering
\includegraphics[width=0.7\textwidth]{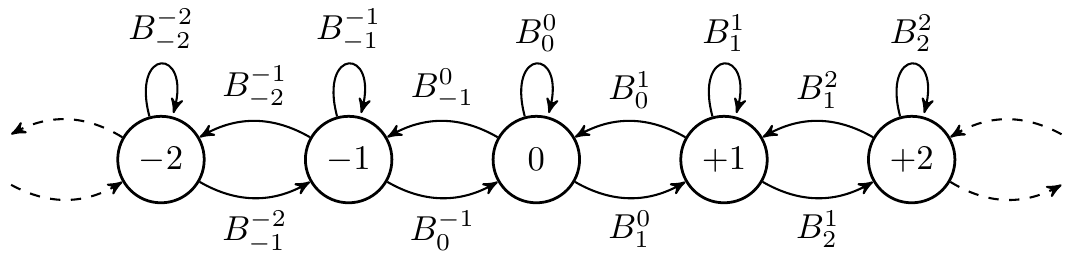}
\caption{The above figure is an illustration of the OQW on $\mathbb{Z}$. The operators $B_j^i$ represent the transition of the ``walker'' jumping from node ($i$) to node ($j$).}
\label{fig:example}
\end{figure}

Now, to describe the dynamics of the walker for each edge $(i, j)$ we introduce bounded operators $B_j^i$ on $\mathcal{H}_S$. These operators represents the change in the internal degree of freedom of the walker due to the effect of passing from vertex $i$ to vertex $j$  (see Fig. \ref{fig:example}). By assuming that for each node $j$,

\begin{equation} \label{eq:2y}
 \sum_i B_j^{i \dag} B_j^i = I,
\end{equation}

\noindent
we make sure that for each node $j\in \mathcal{V}$ there is a corresponding CPTP map $\mathcal{M}_j$ in the Kraus representation on the space of operators on $\mathcal{H}_S$,

\begin{equation} \label{eq:3y}
\mathcal{M}_j (\tau) = \sum_i B_j^i \tau B_j^{i \dag}.
\end{equation}

\noindent
The transition operators $B_j^i$ act only on the internal state Hilbert space $\mathcal{H}_S$ and do not perform transitions from node $i$ to node $j$. They can be easily dilated to operators $M_j^i$ acting on the total Hilbert space $\mathcal{H}_S\otimes \mathcal{K}$ as $M_j^i = B_j^i \otimes \ket{i} \bra{j}$. Hence, it is clear that if condition  (\ref{eq:2y}) is satisfied, then $\sum_{i,j} M_j^{i \dag} M_j^i = I$ \cite{attal2012open}. This condition defines a CPTP map for density matrices on $\mathcal{H}_S \otimes \mathcal{K}$, i.e., 

\begin{equation} \label{eq:4y}
 \mathcal{M} (\rho) = \sum_i \sum_j M_j^i \rho M_j^{i \dag}.
\end{equation}
\noindent
At this point it is worth mentioning that the CPTP map  $\mathcal{M}$ defines the discrete-time OQW \cite{attal2012open, attal2012oopen}. As an example, lets suppose that the state of the walker at any time-step is given by,

\begin{equation} \label{eq:1y}
\rho^{[t]} = \sum_k \rho_k^{[t]} \otimes \ket{k}\bra{k},
\end{equation}

\noindent
where $\rho_k^{[t]}$ is the positive operator on the Hilbert space of the walker describing the state of the inner degree of freedom of the walker at lattice site $k$, satisfying the probability conservation condition $\sum_k \text{Tr}\rho_k^{[t]} $=1. The superscript $[t]$ denotes the time-step $t$.  The basis vector in a Hilbert space corresponding to the graph of the OQW is given by $\ket{k}$. The OQW is obtained by the iteration of the CPTP map as in (\ref{eq:4y}). For the OQWs on the line, the general form of the iteration is given by, 

\begin{eqnarray}\label{eq:yx1}
 \rho_{k}^{[t+1]}&=& B_k^k\rho_{k}^{[t]}B_k^{k \dag}+B_{k-1}^k\rho_{k-1}^{[t]}B_{k-1}^{k\dag}+B_{k+1}^k\rho_{k+1}^{[t]}B_{k+1}^{k\dag}.
\end{eqnarray}

\noindent
The schematics of an OQW on $\mathbb{Z}$ is given by Figure \ref{fig:example}. The completely positivity and trace preservation of the map (\ref{eq:yx1}) is guaranteed by the normalization condition,

\begin{eqnarray}\label{eq:yxh1}
\forall k, \ \ \ B_k^{k \dag}B_k^k + B_k^{k+1 \dag}B_k^{k+1}+B_k^{k-1\dag}B_k^{k-1} = I.
\end{eqnarray}

\noindent
Equation (\ref{eq:yxh1}) denotes the conservation of the probability and is related to a generalization of the classical Markov chain approach in the quantum domain \cite{gudder2008quantum, gudder2010effect}. 

\section{Model}\label{model}


The open quantum system that we consider here was initially proposed by \cite{sinayskiy2014quantum} using a dissipative out-of-resonance cavity quantum electrodynamics setup. In this scheme, the Fock states of the cavity mode correspond to the nodes of the ``walker'', and the state of the two-level system corresponds to an inner degree of freedom of the ``walker''. In this paper we propose a generalized version of the previous scheme by including the temperature of the environment. By introducing the above, the dynamics of the system is now governed by the quantum master equation of the following form \cite{breuer2002theory, carmichael2002, gardiner2004quantum}: 

\begin{equation} \label{eq:1}
 \frac{d}{dt}  \rho = -\imath \left[ H_{\mathrm{int}}, \rho\right] + \gamma \left(\mathrm{n}_\mathrm{th}+1\right)\pazocal{L} \left[ \sigma_{-}, \sigma_{+}\right]\rho + \gamma \mathrm{n}_\mathrm{th} \pazocal{L} \left[ \sigma_{+}, \sigma_{-}\right]\rho,
\end{equation}

\noindent
where $\rho$ is density operator, $H_{\mathrm{int}}$ is the Hamiltonian that describes the interaction between the cavity mode and the two-level atom, $\gamma$ is the spontaneous emission rate, $\mathrm{n}_\mathrm{th} = (e^{\beta \hbar \omega}-1)^{-1}$ is the average thermal photon number, where $\beta$ is the inverse temperature of the environment $\beta = (k_B T)^{-1}$, and $k_B$ is the Boltzmann’s constant. $\pazocal{L}$ denotes the standard dissipative superoperator in the Gorini-Kossakowski-Sudarshan Lindblad (GKSL) 
form \cite{gorini1976completely,lindblad1976generators}. The atom-field interaction Hamiltonian within the rotating-wave approximation (RWA) is given by (in units of $\hbar$ = 1), 

\begin{equation}\label{eq:2}
H_{\mathrm{int}} = \Delta a^{\dag}a + g\left(a\sigma_{+} + a^{\dag}\sigma_{-} \right).
\end{equation}

\noindent
In the Eq. (\ref{eq:2}) $\Delta = \omega_f - \omega_a$ is the detuning between the cavity mode and transition frequency in a two-level atom. Also, $\omega_f$ is the frequency of the cavity field
and $\omega_a$ is the atomic transition frequency. The operators, $a$ and $a^{\dag}$ are the corresponding photon annihilation and creation operators of the electromagnetic field, respectively, satisfying commutation relations $\left[a,a^{\dag}\right] = 1$. The constant $g$ stands for the atom-field coupling strength, which was introduced by Jaynes and Cummings \cite{jaynes1963comparison} to study the interaction between a two-level atom and a quantized single mode field. The Pauli raising and lowering spin operators are denoted by $\sigma_{\pm}$, where $\sigma_{+} = \ket{e}\bra{g}$, and $\sigma_{-} = \ket{g}\bra{e}$, $\ket{e}$ being the excited state and $\ket{g}$ the ground state of the two-level atom. These operators $\sigma_+$, $\sigma_-$ and $\sigma_z$ (generators of $\mathit{su}$(2) group) obey the commutation relations, $\left[\sigma_z,\sigma_{\pm}\right] = \pm 2 \sigma_{\pm}$ and $\left[\sigma_+,\sigma_-\right] = \sigma_z$. In Eq. (\ref{eq:1}) the superoperator $\pazocal{L} \left[ \sigma_{-}, \sigma_{+}\right]\rho$ together with `1' term describes the spontaneous emission of the two-level atom, and the $\mathrm{n}_\mathrm{th}$ one is due to the stimulated process. The last term in (\ref{eq:1}) describes the process in which the two-level atom absorbs photons from the field. The action of the superoperator $\pazocal{L} \left[ n, m\right]$ is defined as:

\begin{equation}\label{eq:3}
\pazocal{L} \left[ n, m\right]\rho = n\rho m - \frac{1}{2}\{mn,\rho \},
\end{equation}

\noindent
where $\{\cdot,\cdot\}$ stands for the anti-commutator. Despite its simplicity, the Hamiltonian (\ref{eq:2}) can not be diagonalized exactly. This will be further discussed in the next few lines. Also, at this point, it is worth mentioning that the dynamics described by Eq. (\ref{eq:1}) is not yet in the OQWs form. 
In order to generate the effective dynamics we move our system into the dispersive media by considering the limit where $\epsilon = \frac{g}{\Delta}\ll 1$. 
The main purpose of this step is to obtain the effective Hamiltonian and effective master equation using the method of small unitary rotations \cite{klimov2000method,klimov2009group,Klimov_2002}.
This technique allows for the approximate diagonalization of nonlinear Hamiltonians. To illustrate this method, we consider some physical system whose interaction Hamiltonian is given by,

\begin{equation} \label{eq:211}
H_{\mathrm{int}} = \Delta A_3 + g\left( A_+ +  A_- \right).
\end{equation}
\noindent
Here also $g$ is a coupling constant, $\Delta$ is a detuning between frequencies of different subsystems. The operators $A_{\pm}$ and $A_3$ satisfy the following commutation relations,

\begin{equation} \label{eq:219}
\left[ A_3, A_{\pm} \right] =  \pm A_{\pm}, \hspace{4 mm} \text{and} \hspace{4 mm} \left[ A_+, A_- \right] = \mathrm{P}\left(A_3, N_i  \right).
\end{equation}

\noindent
In Eq. (\ref{eq:219}) P(A$_3, N_i$) refers to a polynomial function of the diagonal operator $A_3$ with coefficients depending on some integrals of motion $N_i$. These commutation relations correspond to a polynomial of deformation of $\mathit{su}$(2) \cite{klimov2000method,klimov2009group,Klimov_2002}. As mentioned earlier, we consider a physical system where for some physical reason, the ratio between the coupling constant $g$ and the detuning $\Delta$ is a small parameter $\epsilon = \frac{g}{\Delta}\ll 1$. In this case the interaction Hamiltonian $H_\mathrm{int}$ can be approximately diagonalized by applying the small unitary rotation operator $U = \mathrm{exp}\left[\epsilon \left(A_+ -  A_-  \right) \right]$ as $H_{\mathrm{eff}} = U H_{\mathrm{int}} U^{\dag}$: hence, the master equation transforms as $\rho_{\mathrm{eff}} = U \rho U^{\dag}$. Using the usual expansion $e^A B e^{-A}=B+[A, B]+\frac{1}{2!}[A,[A,B]]+ \cdot \cdot \cdot$ and keeping terms up to order $\epsilon^2$, one gets the Hamiltonian that is diagonal in the basis of the operator $A_3$, defined by,

 \begin{equation} \label{eq:913}
H_{\mathrm{eff}} = \Delta A_{3} + \frac{g^2}{\Delta} \mathrm{P}\left(A_{3},N_{i}  \right).
\end{equation}

\noindent
For the problem considered in this paper the appropriate operator of the small unitary rotation has the following form,
 \begin{equation} \label{eq:914}
U = \mathrm{exp}\left[\epsilon \left(  a^{\dag} \sigma_- -  a \sigma_+  \right) \right].
\end{equation}

\noindent
After some calculations, we obtain the effective master equation of the following form,

\begin{eqnarray}\label{eq:131}
\frac{d }{dt} \rho_{\mathrm{eff}}& = & -\imath \left[ H_{\mathrm{eff}}, \rho_{\mathrm{eff}}\right] + \gamma \left(\mathrm{n}_\mathrm{th} +1 \right)  \biggl[\pazocal{L} \left[ \sigma_-, \sigma_+\right] \rho_{\mathrm{eff}} +\frac{g}{\Delta}\pazocal{L} \left[a \sigma_{z}, \sigma_+\right] \rho_{\mathrm{eff}} + \frac{g}{\Delta}\pazocal{L} \left[\sigma_-, a^{\dag}\sigma_{z}\right] \rho_{\mathrm{eff}}\nonumber \\ \nonumber
& & +  \frac{g^2}{{\Delta}^2}\pazocal{L} \left[a \sigma_{z}, a^{\dag} \sigma_{z} \right] \rho_{\mathrm{eff}} -  \frac{g^2}{{\Delta}^2}\pazocal{L} \left[ \sigma_-, \left(a^{\dag} a + 1\right) \sigma_+ + 2 a^{{\dag}^2}\sigma_- \right] \rho_{\mathrm{eff}} \\ \nonumber
& & - \frac{g^2}{{\Delta}^2}\pazocal{L} \left[ \left(a^{\dag} a + 1\right) \sigma_- + 2 a^{{\dag}^2}\sigma_+,  \sigma_+ \right] \rho_{\mathrm{eff}}\biggl] + \gamma \mathrm{n}_\mathrm{th} \biggl[ \pazocal{L} \left[ \sigma_+, \sigma_-\right] \rho_{\mathrm{eff}} -\frac{g}{\Delta}\pazocal{L} \left[\sigma_+, a \sigma_z\right] \rho_{\mathrm{eff}}\\ \nonumber
& & - \frac{g}{\Delta}\pazocal{L} \left[a^{\dag}\sigma_z, \sigma_-\right] \rho_{\mathrm{eff}}+  \frac{g^2}{{\Delta}^2}\pazocal{L} \left[a^{\dag} \sigma_z, a \sigma_z \right] \rho_{\mathrm{eff}} -  \frac{g^2}{{\Delta}^2}\pazocal{L} \left[ \sigma_+, \left(a^{\dag} a + 1\right) \sigma_- + 2 a^2\sigma_+ \right] \rho_{\mathrm{eff}} \\
& & - \frac{g^2}{{\Delta}^2}\pazocal{L} \left[ \left(a^{\dag} a + 1\right) \sigma_+ + 2 a^{{\dag}^2}\sigma_-,  \sigma_- \right] \rho_{\mathrm{eff}}  -\frac{g^2}{{\Delta}^2}\pazocal{L} \left[\sigma_+,  \sigma_- \right] \rho_{\mathrm{eff}}\biggl],
\end{eqnarray}

\noindent
where the effective Hamiltonian $H_\mathrm{eff}$ reads,

\begin{equation} \label{eq:11}
H_{\mathrm{eff}} = \Delta a^{\dag}a   - \frac{g^2}{\Delta}\left( a^{\dag}a \sigma_{z} + \frac{\sigma_{z}}{2} + \frac{1}{2}  \right).
\end{equation}

\noindent
From (\ref{eq:11}), the diagonal operator $A_3$ corresponds to $a^{\dag}a$. The total excitation number $N=a^{\dag}a + \sigma_z$ is a constant of motion. All terms in the master equation (\ref{eq:131}), which depend on non-equal powers of the photon annihilation and creation operators $a$ and $a^{\dag}$, will oscillate rapidly and do not contribute to the system dynamics. After performing the rotating wave approximation (RWA) the effective master equation takes the form,

\begin{eqnarray} \label{eq:1w1}
\frac{d }{dt} \rho_{\mathrm{eff}}& = & \imath \frac{ g^2}{\Delta} \left[a^{\dag}a\sigma_{z}  + \frac{\sigma_{z}}{2} +\frac{1}{2}, \rho_{\mathrm{eff}}\right]  + \gamma \left(\mathrm{n}_\mathrm{th} +1 \right)  \biggl[\pazocal{L} \left[ \sigma_-, \sigma_+\right] \rho_{\mathrm{eff}}+  \frac{g^2}{{\Delta}^2}\pazocal{L} \left[a \sigma_{z}, a^{\dag} \sigma_{z} \right] \rho_{\mathrm{eff}} \nonumber \\ \nonumber
& &  -  \frac{g^2}{{\Delta}^2}\pazocal{L} \left[ \sigma_-, \left(a^{\dag} a + 1\right) \sigma_+ \right] \rho_{\mathrm{eff}} - \frac{g^2}{{\Delta}^2}\pazocal{L} \left[ \left(a^{\dag} a + 1\right) \sigma_-,  \sigma_+ \right] \rho_{\mathrm{eff}}\biggl] \\ \nonumber
& & + \gamma \mathrm{n}_\mathrm{th} \biggl[ \pazocal{L} \left[ \sigma_+, \sigma_-\right]\rho_{\mathrm{eff}}+ \frac{g^2}{{\Delta}^2}\pazocal{L} \left[a^{\dag} \sigma_z, a \sigma_z \right] \rho_{\mathrm{eff}} -  \frac{g^2}{{\Delta}^2}\pazocal{L} \left[ \sigma_+, \left(a^{\dag} a + 1\right) \sigma_- \right] \rho_{\mathrm{eff}} \\ 
& & - \frac{g^2}{{\Delta}^2}\pazocal{L} \left[ \left(a^{\dag} a + 1\right) \sigma_+,  \sigma_- \right] \rho_{\mathrm{eff}}  -\frac{g^2}{{\Delta}^2}\pazocal{L} \left[\sigma_+,  \sigma_- \right] \rho_{\mathrm{eff}}\biggl].
\end{eqnarray}

\noindent
From this point, one can see that the above master equation (\ref{eq:1w1}) conserves the typical form of the OQWs introduced earlier (\ref{eq:1y}). By writing the density matrix of the reduced system from Eq. (\ref{eq:1w1}), as $\rho = \sum_k \rho_k \otimes \ket{k}\bra{k}$, where $\ket{k}$ is a Fock state of the cavity mode and $\rho_k$ is a positive operator describing the state of the two-level system, the quantum master equation (\ref{eq:1w1}) reduces to the system of differential equations for the operators $\rho_k$:

\begin{eqnarray}\label{eq:119}
 \frac{d}{dt}  \rho_{k}&=& \imath \frac{g^2}{\Delta}\left[k \sigma_{z}+\frac{\sigma_{z}}{2} +\frac{1}{2},   \rho_{\mathrm{k}} \right]  +  \gamma \left(\mathrm{n}_\mathrm{th} +1 \right)\frac{g^2}{{\Delta}^2}\biggl[\left(k+1\right)\sigma_z \rho_{\mathrm{k+1}}\sigma_z - k \rho_{\mathrm{k}}\biggl]\nonumber \\ \nonumber
 & &+   \gamma \left(\mathrm{n}_\mathrm{th} +1 \right)\biggl[1 - \frac{2g^2}{{\Delta}^2}\left(k+1\right)\biggl]\pazocal{L} \left[ \sigma_-, \sigma_+\right] \rho_{\mathrm{k}}\\
 & & +  \gamma \mathrm{n}_\mathrm{th} \biggl[(1 - \frac{g^2}{{\Delta}^2}\left(2k+3\right)) \pazocal{L} \left[ \sigma_+, \sigma_-\right] \rho_{\mathrm{k}}+\frac{g^2}{{\Delta}^2}\left(k\sigma_z \rho_{\mathrm{k-1}}\sigma_z -\left(k+1 \right) \rho_{\mathrm{k}}\right)\biggl].
\end{eqnarray}

\noindent
The system of differential equations (\ref{eq:119}) defines the continuous-time OQWs \cite{pellegrini2014continuous}. In order to obtain a discrete-time OQW in the form (\ref{eq:4y}), we construct the explicit form of the transition operators $B_j^i$ by introducing discretized time steps. To do this, we discretize the system of differential equations by replacing the time derivatives by the finite difference with a small time step $\delta t$,

\begin{equation} \label{eq:21}
\frac{d}{dt}\rho_k (t) \rightarrow \frac{\rho_k (t+\delta t)-\rho_k(t)}{\delta t}.
\end{equation}

\noindent
The above substitution leads to the following  jump operators:

\begin{align}\label{eq:xp1}
B_k^k=&\sqrt{\gamma (\mathrm{n}_\mathrm{th}+1)\delta t\left(1-\frac{2g^2}{\Delta^2}(k+1)\right)}\sigma_-, \hspace{3 mm} B_{k+1}^k=\sqrt{\frac{\gamma (\mathrm{n}_\mathrm{th}+1)g^2 \delta t (k+1) }{{\Delta}^2}}\sigma_z,            \nonumber  \\ 
{B^{\prime}}_k^k=&\sqrt{\gamma \mathrm{n}_\mathrm{th}\delta t\left(1-\frac{g^2}{\Delta^2}(2k+3)\right)}\sigma_+, \hspace{3 mm}        B_{k-1}^k=\sqrt{\frac{\gamma \mathrm{n}_\mathrm{th} g^2 \delta t k}{{\Delta}^2}}\sigma_z,& \\ \nonumber 
{B^{\prime \prime}}_k^k=& 1 + \imath \frac{g^2 \delta t}{\Delta} \left(k \sigma_{z}+\frac{\sigma_{z}}{2} +\frac{1}{2}\right)-\frac{\gamma (\mathrm{n}_\mathrm{th}+1)g^2 \delta t k }{2{\Delta}^2}- \frac{1}{2}\gamma (\mathrm{n}_\mathrm{th}+1)\delta t\left(1-\frac{2g^2}{\Delta^2}(k+1)\right)\sigma_+ \sigma_- \\
&-\frac{\gamma \mathrm{n}_\mathrm{th}\delta t}{2} \left(1-\frac{g^2}{\Delta^2}(2k+3)\right)\sigma_-\sigma_+-\frac{\gamma \mathrm{n}_\mathrm{th} g^2 \delta t (k+1)}{2\Delta^2}. \nonumber
\end{align}

\noindent
Hence, the iteration formula for the discrete-time OQW is given by,
\begin{eqnarray}\label{eq:x1}
 \rho_{k}^{[t+1]}&=& B_k^k\rho_{k}^{[t]}B_k^{k \dag}+{B^{\prime}}_k^k\rho_{k}^{[t]}{B^{\prime}}_k^{k \dag}+{B^{\prime \prime}}_k^k\rho_{k}^{[t]}{B^{\prime \prime}}_k^{k \dag}+B_{k-1}^k\rho_{k-1}^{[t]}B_{k-1}^{k\dag}+B_{k+1}^k\rho_{k+1}^{[t]}B_{k+1}^{k\dag}.
\end{eqnarray}

\begin{figure}[H]
   \centering
   \subfigure[]{\includegraphics[width=.47\textwidth]{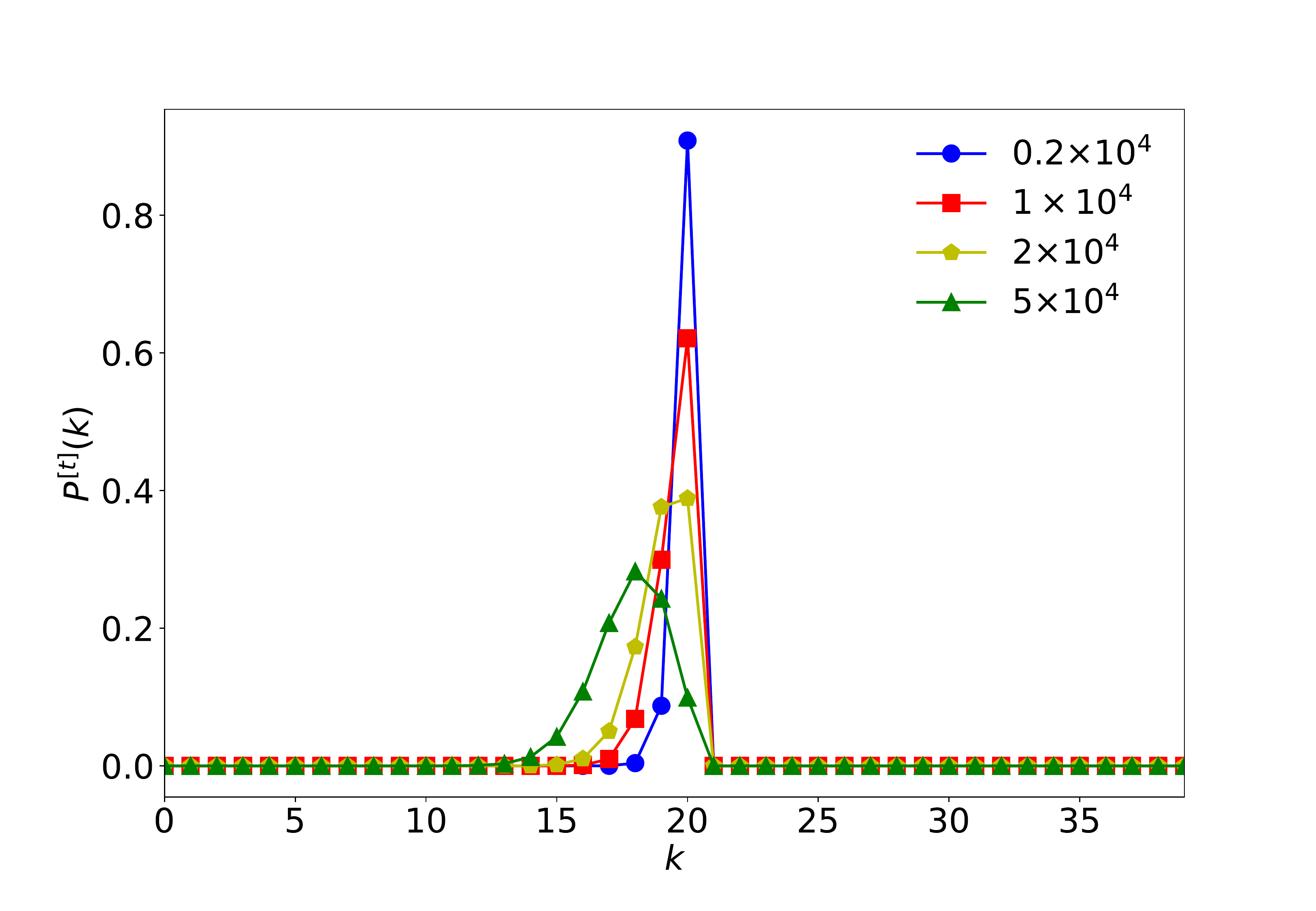}}
   \qquad
   \subfigure[]{\includegraphics[width=.47\textwidth]{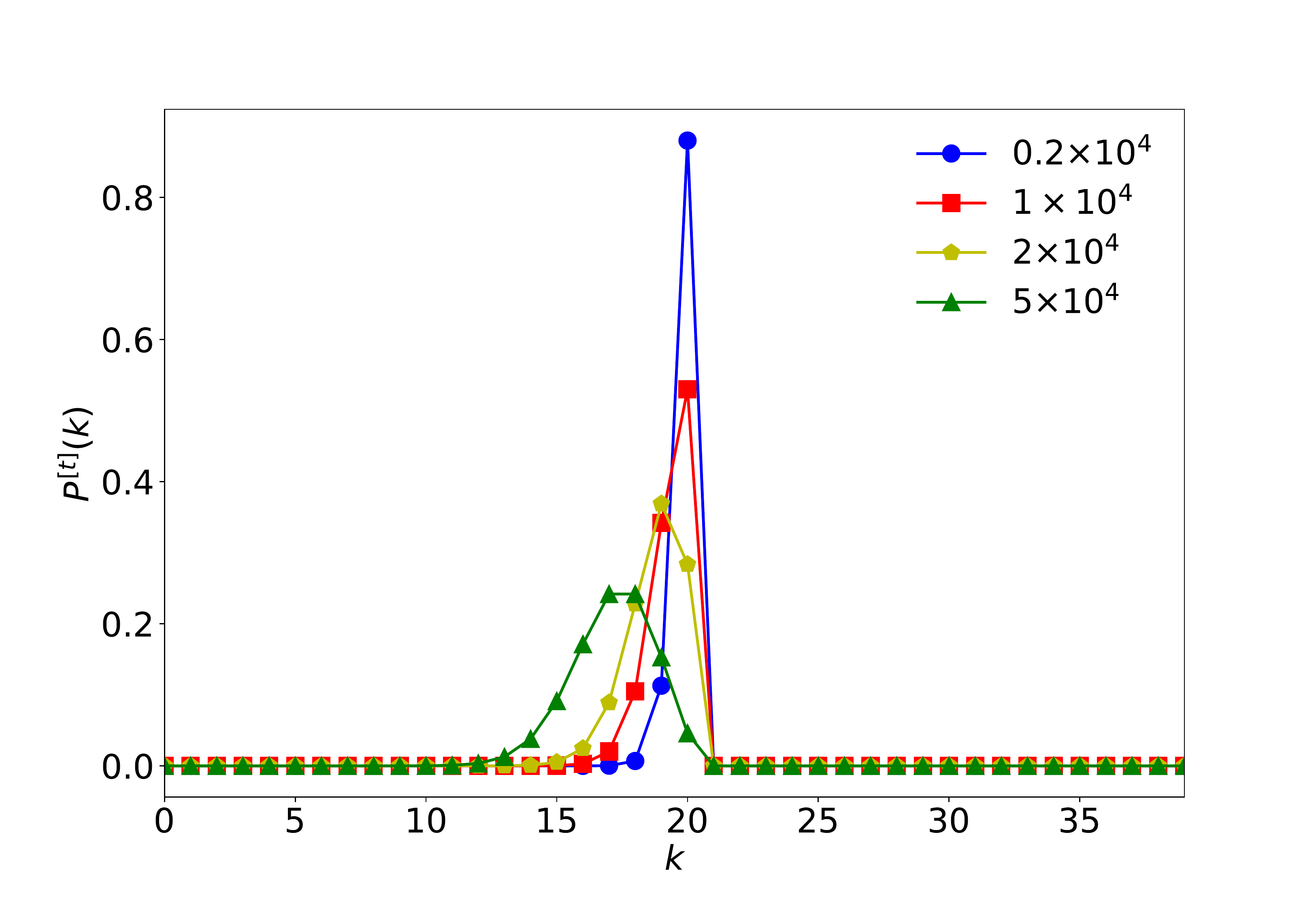}}
   \subfigure[]{\includegraphics[width=.47\textwidth]{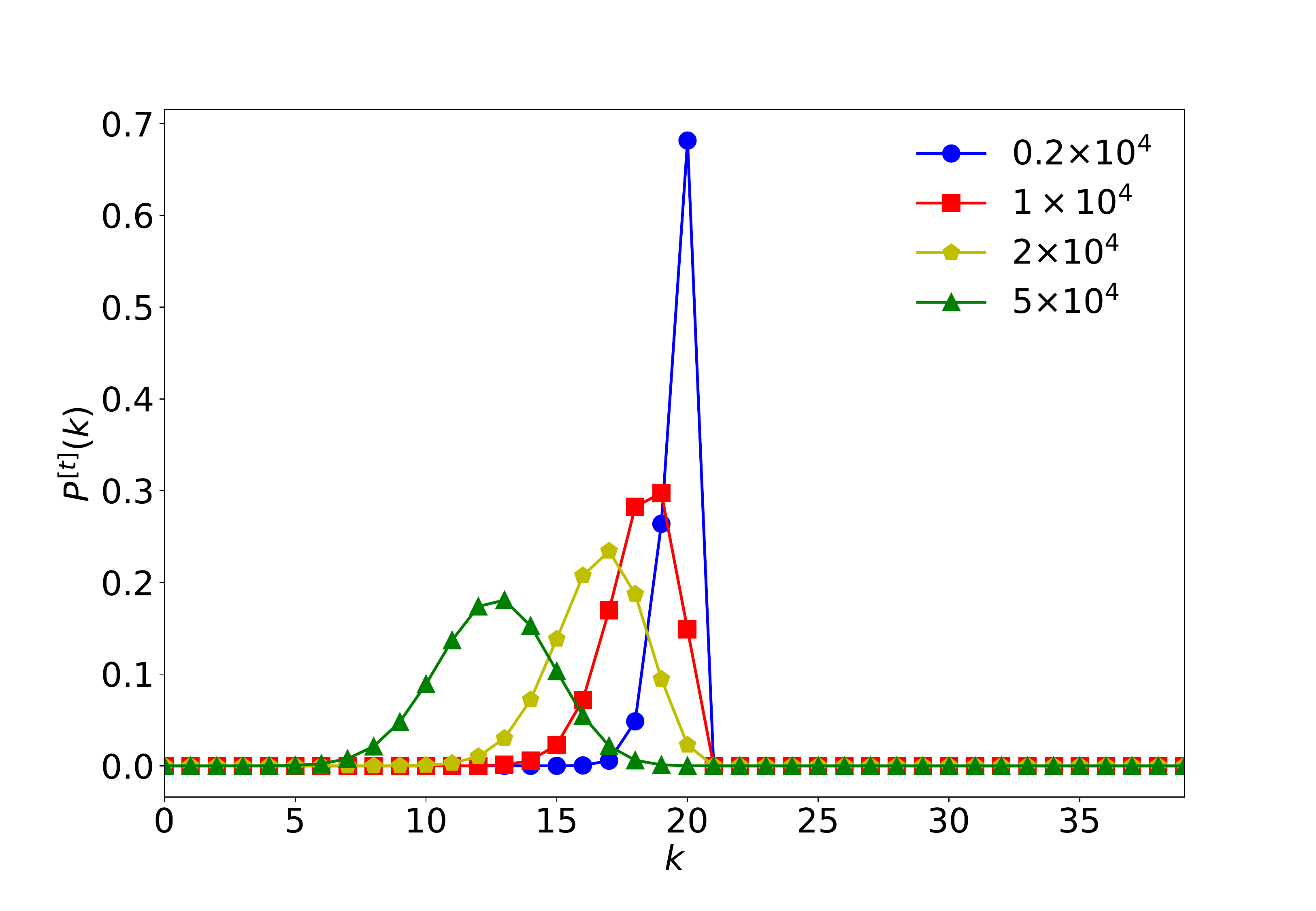}}
   \qquad
   \subfigure[]{\includegraphics[width=.47\textwidth]{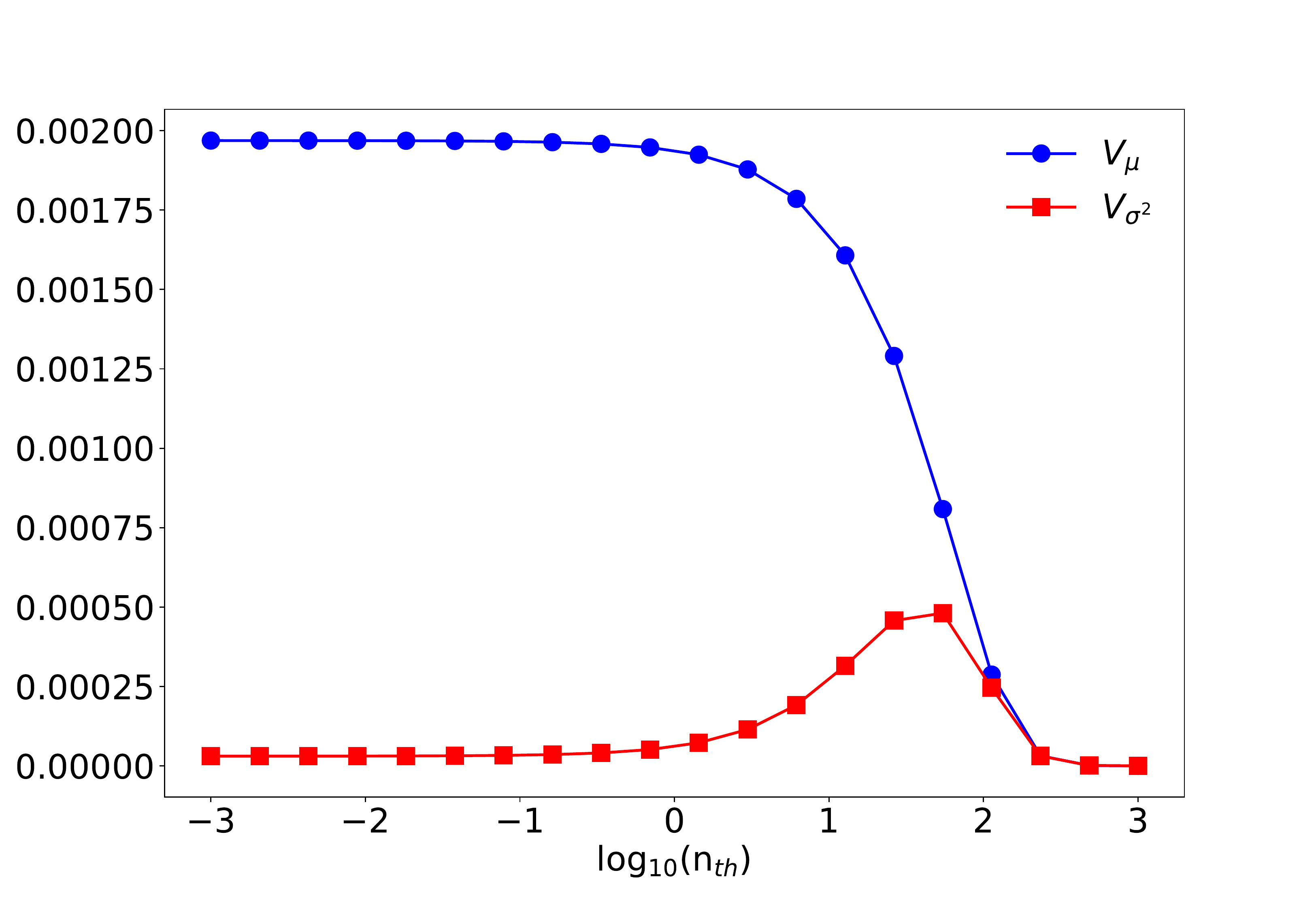}}
   \caption{\small \small (Color online) Open quantum walk (OQW) observables. Figure (a), (b) and (c) shows the occupation probability distribution for the ``walker'' in the generalized quantum optical implementation of OQW at different temperatures of the environment, $\mathrm{n}_{\mathrm{th}}$ = 0.5, 1, and 5, respectively; with iteration formula given by Eq. (\ref{eq:x1}). The markers, namely, circles, squares, pentagon and triangles corresponds to probability distribution after 0.2$\times 10^4$, $ 10^4$, and 5$\times 10^4$ steps. Figure (d) shows the dependence of the ``speed'' $V_\mu$ and ``spread'' $V_\sigma$ of the Gaussian given by Eq. (\ref{eq:11xgh}) and Eq. (\ref{eq:11xgt}) as function temperature  of the environment after $10^4$ steps. The initial state is ${\rho}^{[0]} = \ket{0}\bra{0}\otimes\ket{20}\bra{20}$, where $\ket{20}$ is the initial 20-photon Fock state used as initial state of the cavity mode, which corresponds to an initial lattice site for the OQW. Other parameters used were, $g=0.02$, $\Delta=1$, $\gamma=0.2$, and $\delta t=0.02$.}
   \label{fig:sub1}
\end{figure}

\noindent
The generalized scheme shows that the OQW is driven by two jump operators, for the left and right movements, namely, $B_{k-1}^k$ (to the left) and $B_{k+1}^k$ (to the right). And also the other three jump operators which act on sites $B_k^k$, ${B^\prime}_k^k$ and ${B^{\prime \prime}}_k^k$ do not contribute to the either movements. Hence, this seems to agree very well with the iteration formula (\ref{eq:yx1}). One can easily verify the normalization condition  (\ref{eq:yxh1}) by dropping terms of order ${\delta t}^2$ in Eq. (\ref{eq:xp1}). The jump operators $B_j^i$  (\ref{eq:xp1}) depend on the number of the nodes $k$ which defines an inhomogeneous OQW on the line \cite{sinayskiy2014quantum}. As mentioned earlier, the set of nodes for the walk corresponds to the different Fock states and jumps between the different nodes correspond to the action of the annihilation and creation operator on the Fock states. Figure \ref{fig:sub1} shows the dynamics of different observables for the OQW.  The occupation probability distribution for the ``walker'' $P^{[t]} (k) =$ Tr$\left[\rho_k^{[t]}\right]$ for different numbers of time steps is shown in Fig. \ref{fig:sub1}(a), (b) and (c).

\noindent
One can see that increasing the temperature of the environment (from Fig. \ref{fig:sub1}(a) to Fig. \ref{fig:sub1}(c)) the Gaussian distribution describing the occupation probability of the position of the ``walker'' moves faster to the left and in very low temperatures of the environment the average position of ``walker'' remains near the initial node 20. Also, it is very clear that for the parameters chosen, it takes at least 5$\times 10^4$ steps (Fig. \ref{fig:sub1}(a), triangles) for the ``walker'' to reach the asymptotic distribution. But this is strongly dependent on the temperature of the environment, as can be seen from Fig. \ref{fig:sub1}(c), the asymptotic distribution is reached much earlier at 2$\times 10^4$ steps. As we increase the number of time steps (for 2$\times 10^4$ and  5$\times 10^4$ in Fig. \ref{fig:sub1}(c)) at higher temperatures the width of the Gaussian distributions seems to be the same. In order to understand this, one needs to analyse the mean $\mu (t)$ and variance $\sigma^2 (t)$ of the position of the ``walker'' using the following equations,

\begin{equation} \label{eq:11xgh}
V_\mu = \frac{\mu (t)}{t} \hspace{0.2 cm} \text{where} \hspace{0.2 cm} \mu (t) = \sum_k k P_k^{[t]},
\end{equation}

and,

\begin{equation} \label{eq:11xgt}
V_{{\sigma}^2} = \frac{\sigma^2}{t} \hspace{0.2 cm} \text{where} \hspace{0.2 cm} \sigma^2 (t) = \sum_k {\left(k-\mu (t) \right)}^2P_k^{[t]}.
\end{equation}

\noindent
Fig. \ref{fig:sub1}(d) shows the dependence of the ``speed'' $V_\mu$ and ``spread'' $V_\sigma$ of the Gaussian for different temperatures of the environment on a logarithmic scale. This figure illustrate the dynamics of the Gaussians corresponding to  $N_{\text{steps}} = 10000$ steps shown on Fig. \ref{fig:sub1}(a) to Fig. \ref{fig:sub1}(c). One can see that the biggest change in the velocity is happening for the temperature corresponding to the average number of photons in the environment between 3 (log${_{10}}$(n$_{\mathrm{th}}$) = 0.5) and 100 (log${_{10}}$(n$_{\mathrm{th}}$) = 2). The dependence of the ``spread'' of the Gaussians as a function of temperature (Fig. \ref{fig:sub1}(d)) is sharply decreasing. A similar behavior has been reported in the previous study by \cite{sinayskiy2015microscopic} where increasing the temperature of the environment, causes the ``spread'' to grow to a certain point and decreases afterward.

\section{Conclusion}\label{Conclusion}

In this paper, we proposed a generalized quantum optical implementation of the OQW. This was done by including non-zero temperature of the environment to the scheme suggested earlier by \cite{sinayskiy2014quantum}. The master equation describing the effective dynamics of the system was derived using the method of small unitary rotation approach. However, the resulting master equation defines a continuous time OQW. This master equation conserves the diagonal in position form of the reduced density matrix. A discrete-time version corresponding to the CPTP maps was obtained using a discretization procedure.

In summary, we demonstrated that the ``walker'' reaches the asymptotic distribution. The temperature of the environment plays an important role in OQWs because it allows the system to reach a steady Gaussian distribution faster. Future work will show how to implement a more diverse OQW by using a microscopic maser setup \cite{gleyzes2007quantum,brune1992manipulation,guerlin2007progressive}. This scheme will be based  on a quantum-non-demolition method to measure the number of photons stored in a high-Q cavity.

\section*{Acknowledgements}

This work is based upon research supported by the South African Research Chair Initiative of the Department of Science and Technology and National Research Foundation. AZ acknowledge support in part by the National Research Foundation of South Africa (Grant No. 118892).

\nocite{*}
\bibliographystyle{unsrt}
\bibliography{generic}
\end{document}